%% file: dmc-main.tex
\documentclass[11pt]{llncs}
\usepackage[utf8]{inputenc}

\input{macros}

\title{Specifying and Verifying Properties of Space \\ Extended version\thanks{Research partially funded by EU 
ASCENS (nr. 257414), EU QUANTICOL (nr. 600708), 
IT MIUR CINA and PAR FAS 2007-2013 Regione Toscana  TRACE-IT.}}
\author{Vincenzo Ciancia\inst{1}\and Diego Latella\inst{1}\and Michele Loreti\inst{2}\and Mieke Massink\inst{1}}

\institute{Istituto di Scienza e Tecnologie dell'Informazione \lq A.~Faedo\rq, CNR, Italy \and
Universit\`a di Firenze, Italy
}

\newcommand{\slcs}{SLCS}

\begin{document}

\maketitle

\begin{abstract}
    The interplay between process behaviour and spatial aspects of computation has become more and more relevant in Computer Science, especially in the field of \emph{collective adaptive systems}, but also, more generally, when dealing with systems distributed in physical space. Traditional verification techniques are well suited to analyse the temporal evolution of programs; properties of space are typically not explicitly taken into account. We propose a methodology to verify properties depending upon physical space. We define an appropriate logic, stemming from the tradition of topological interpretations of modal logics, dating back to earlier logicians such as Tarski, where modalities describe neighbourhood. We lift the topological definitions to a more general setting, also encompassing discrete, graph-based structures. We further extend the framework with a spatial \emph{until} operator, and define an efficient model checking procedure, implemented in a proof-of-concept tool. 
\end{abstract}



\input{introduction.tex}

\input{closure-spaces.tex}


\input{quasi-discrete.tex}


\input{digital-spatial-logic.tex}

\input{model-checking.tex}
\input{examples.tex}


\input{future-work.tex}

\bibliographystyle{plain}
\bibliography{bibliography.bib}

\newpage
\appendix

\input{proofs.tex}

\end{document}

%% file: macros.tex
\usepackage{url}
\usepackage{amssymb}
\usepackage{amsmath}
\usepackage{tikz}
\usepackage{stmaryrd}
\usepackage{graphicx}
\usepackage[noline,noend]{algorithm2e}
\def\algorithmautorefname{Algorithm}

\usepackage{verbatim}

\usepackage{hyperref}
\hypersetup{pdfborder={0 0 0}}

\newcommand{\eval}{\mathcal{V}}

\newcommand{\props}{P}

\newcommand{\model}{\mathcal{M}}
\newcommand{\nmodels}{\nvDash}

\newcommand{\until}{\,\mathcal{U}}

\newcommand{\boundary}{\mathcal{B}}
\newcommand{\iboundary}{\boundary^-}
\newcommand{\cboundary}{\boundary^+}
\newcommand{\closure}{\mathcal{C}}
\newcommand{\interior}{\mathcal{I}}

\newcommand{\nats}{\mathbb{N}}

\newcommand{\ints}{\mathbb{Z}}

\newcommand{\lboundary}{\mathcal{\partial}}
\newcommand{\liboundary}{\lboundary^-}
\newcommand{\lcboundary}{\lboundary^+}

\renewcommand{\succ}{\mathit{Succ}}
\newcommand{\pto}[2]{\overset{#1}{\underset{#2}{\rightsquigarrow}}\infty}
\newcommand{\ldualuntil}{\, \mathcal{R}} 
\newcommand{\leverywhere}{\mathcal{G}}
\newcommand{\lsomewhere}{\mathcal{F}}

\newcommand{\mneigh}{N}

\numberwithin{theorems}{section}

\usepackage{mfirstuc}
\usepackage{aliascnt}

\newcommand{\NewTheorem}[1]{
  \newaliascnt{#1}{theorems}
  \newtheorem{#1}[#1]{\makefirstuc{#1}}
  \aliascntresetthe{#1}
  \expandafter\def\csname #1autorefname\endcsname{\makefirstuc{#1}}
}

\usepackage{comment}
\newif\ifrelease
\releasefalse

\newcommand{\NewDraftTheorem}[1]{\ifrelease\excludecomment{#1}\else
  \newaliascnt{#1}{theorems}
  \newtheorem{#1}[#1]{*\makefirstuc{#1}*}
  \aliascntresetthe{#1}
  \expandafter\def\csname #1autorefname\endcsname{\makefirstuc{#1}}
\fi}

\NewDraftTheorem{todo}
\NewDraftTheorem{later}
\NewDraftTheorem{internal}

\usepackage{tikz}
\usepackage{graphicx}

\usepackage{tabto}
\newcommand{\dertab}{.9cm}
\newcommand{\dervskip}{.2cm}

\newcommand{\der}{$\vskip \dervskip\noindent\tabto{\dertab}$}

\newcommand{\stp}[1]{\\*[\dervskip]#1$\tabto{\dertab}$}
\newcommand{\jstp}[2]{\\*[\dervskip]#1$\tabto{\dertab}$[\,#2\,]\\*[\dervskip]$\tabto{\dertab}$}

\DontPrintSemicolon
\SetKwFunction{check}{Sat}
\SetKwFunction{checkUntil}{CheckUntil}
\SetKwProg{myfun}{Function}{}{}
\SetKwSwitch{Match}{Case}{Other}{Match}{}{case}{otherwise}{~ }{}
\SetKw{Let}{let}
\SetKw{In}{in}
\SetKw{Var}{var}
\SetKw{Conj}{and}

%% file: introduction.tex
\section{Introduction}

Much attention has been devoted in Computer Science to formal verification of process behaviour. Several techniques, such as \emph{run-time monitoring} and \emph{model-checking}, are based on a formal understanding of system requirements through \emph{modal} logics. Such logics typically  have a \emph{temporal} flavour, describing the flow of events along time, and are interpreted in various kinds of transition structures. 

Recently, aspects of computation related to the distribution of systems in physical space have become more relevant. An example is provided by so called \emph{collective adaptive systems}\footnote{See e.g. the web site of the QUANTICOL project: \url{http://www.quanticol.eu}}, typically composed of a large number of interacting objects. Their global behaviour critically depends on interactions which are often local in nature. Locality immediately poses issues of spatial distribution of objects. Abstraction from spatial distribution may sometimes provide insights in the system behaviour, but this is not always the case. For example, consider a bike (or car) sharing system having several parking stations, and featuring twice as many parking slots as there are vehicles in the system. Ignoring the spatial dimension, on average, the probability to find completely full or empty parking stations at an arbitrary station is very low; however, this kind of analysis may be misleading, as in practice some stations are much more popular than others, often depending on nearby points of interest. This leads to quite different probabilities to find stations completely full or empty, depending on the spatial properties of the examined location. In such situations, it is important to be able to predicate over spatial aspects, and eventually find methods to certify that a given formal model of space
satisfies specific requirements in this respect. In Logics, there is quite an amount of literature focused on so called \emph{spatial} logics, that is, a spatial interpretation of modal logics. Dating back to early logicians such as Tarski, modalities may be interpreted using the concept of \emph{neighbourhood} in a topological space. The field of spatial logics is well developed in terms of descriptive languages and computability/complexity aspects. However, the frontier of current research does not yet address verification problems, and in particular, discrete models are still a relatively unexplored field. 


In this paper, we extend the topological semantics of modal logics to \emph{closure spaces}. As we shall discuss in the paper, this choice is motivated by the need to use non-idempotent \emph{closure operators}. A closure space (also called \emph{\v Cech closure space} or \emph{preclosure space} in the literature), is a generalisation of a standard topological space, where idempotence of closure is not required. By this, graphs and topological spaces are treated uniformly, letting the topological and graph-theoretical notions of neighbourhood coincide. We also provide a spatial interpretation of the \emph{until} operator, which is fundamental in the classical temporal setting, arriving at the definition of a logic which is able to describe unbounded areas of space. Intuitively, the spatial until operator describes a situation in which it is not possible to ``escape'' an area of points satisfying a certain property, unless by passing through at least one point that satisfies another given formula. To formalise this intuition, we provide a characterising theorem that relates infinite paths in a closure space and until formulas. 
We introduce a model-checking procedure that is \emph{linear} in the size of the considered space. A prototype implementation of a spatial model-checker has been made available; the tool is able to interpret spatial logics on digital images, providing graphical understanding of the meaning of formulas, and an immediate form of counterexample visualisation.

\paragraph{Related work.} We use the terminology \emph{spatial logics} in the ``topological'' sense; the reader should be warned that in Computer Science literature, spatial logics typically describe situations in which modal operators are interpreted syntactically, against the structure of agents in a process calculus (see \cite{CG00,CC03} for some classical examples). The object of discussion in this research line are operators that quantify e.g., over the parallel sub-components of a system, or the hidden resources of an agent.  Furthermore, logics for graphs have been studied in the context of databases and process calculi (see \cite{CGG02,GL07}, and references), even though the relationship with physical space is  often not made explicit, if considered at all. The influence of space on agents interaction is also considered in the literature on process calculi using \emph{named locations}~\cite{DFP98}.
Variants of spatial logics have also been proposed for the symbolic representation of the contents of images, and, combined with temporal logics, for sequences of images~\cite{DVD95}. The approach is based on a discretisation of the space of the images in rectangular regions and the orthogonal projection of objects and regions onto Cartesian coordinate axes such that their possible intersections can be analysed from different perspectives. It involves two spatial until operators defined on such projections considering spatial shifts of regions along the positive, respectively negative, direction of the coordinate axes
and it is very different from the topological spatial logic approach.
A successful attempt to bring topology and digital imaging together is represented by the field of \emph{digital topology} \cite{Ros79,KR89}. In spite of its name, this area studies digital images using models inspired by topological spaces, but neither generalising nor specialising these structures. Rather recently, closure spaces have been proposed as an alternative foundation of digital imaging by various authors, especially Smyth and Webster \cite{SW07} and Galton \cite{Gal03}; we continue that research line, enhancing it with a logical perspective. Kovalevsky \cite{Kov08} studied alternative axioms for topological spaces in order to recover well-behaved notions of neighbourhood. In the terminology of closure spaces, the outcome is that one may impose closure operators on top of a topology, that do not coincide with topological closure.
The idea of interpreting the until operator in a topological space is briefly discussed in the work by Aiello and van Benthem \cite{A02,vBB07}. We start from their definition, discuss its limitations, and provide a more fine-grained operator, which is interpreted in closure spaces, and has therefore also an interpretation in topological spaces.  
In the specific setting of complex and collective adaptive systems, techniques for efficient approximation have been developed in the form of mean-field / fluid-flow analysis (see \cite{BHLM13} for a tutorial introduction). Recently (see e.g., \cite{CLR09}), the importance of spatial aspects has been recognised and studied in this context. In this work, we aim at paving the way for the inclusion of spatial logics, and their verification procedures, in the framework of mean-field and fluid-flow analysis of collective adaptive systems.



%% file: closure-spaces.tex
\section{Closure spaces}

In this work, we use \emph{closure spaces} to define 
basic concepts of \emph{space}. Below, we recall several definitions, most of which are explained in \cite{Gal03}. 

\begin{definition}\label{def:closure-space}
A \emph{closure space} is a pair $(X,\closure)$ where $X$ is a set, and the \emph{closure operator} $\closure : 2^X \to 2^X$ assigns to each subset of $X$ its \emph{closure}, obeying to the following laws, for all $A,B \subseteq X$:
\begin{enumerate}
 \item \label{def:closure-space:closure-of_emptyset} $\closure(\emptyset) = \emptyset$;
 \item \label{def:closure-space:closure-larger} $A \subseteq \closure(A)$;
 \item \label{def:closure-space:closure-union} $\closure(A \cup B) = \closure(A) \cup \closure(B)$.
\end{enumerate}
\end{definition}

\noindent As a matter of notation, in the following, for $(X,\closure)$ a closure space, and $A \subseteq X$, we let $\overline A = X \setminus A$ be the complement of $A$ in $X$.

\begin{definition}\label{def:closure-concepts}
Let $(X,\closure)$ be a closure space, for each $A \subseteq X$:
\begin{enumerate}
\item the \emph{interior} $\interior(A)$ of $A$ is the set $\overline{\closure(\overline A)}$;
\item \label{def:closure-concepts:neighbourhood}$A$ is a \emph{neighbourhood} of $x \in X$ if and only if $x \in \interior(A)$;
\item \label{def:closure-concepts:closed-open}$A$ is \emph{closed} if $A = \closure(A)$ while it is \emph{open} if $A = \interior(A)$.
\end{enumerate} 
\end{definition}
 
\begin{lemma}\label{lem:interior-monotone}
Let $(X,\closure)$ be a closure space, the following properties hold:
\begin{enumerate}
\item \label{lem:interior-monotone-1} $A\subseteq X$ is open if and only if $\overline A$ is closed;
\item \label{lem:interior-monotone-2} closure and interior are monotone operators over the inclusion order, that is:
$A \subseteq B \implies \closure(A) \subseteq \closure(B)\mbox{ and }\interior(A) \subseteq \interior(B)$
\item \label{lem:interior-monotone-3} Finite intersections and arbitrary unions of open sets are open.
 \end{enumerate}
\end{lemma}

%

%
%

Closure spaces are a generalisation of \emph{topological spaces}. The axioms defining a closure space are also part of the definition of a \emph{Kuratowski closure space}, which is one of the possible alternative definitions of a topological space. More precisely, a closure space is Kuratowski, therefore a topological space, whenever closure is \emph{idempotent},  
that is, $\closure(\closure(A)) = \closure(A)$. We omit the details for space reasons (see e.g., \cite{Gal03} for more information). 

Next, we introduce the topological notion of \emph{boundary}, which also applies to closure spaces, and two of its variants, namely the \emph{interior} and \emph{closure} boundary (the latter is sometimes called \emph{frontier}).

\begin{definition}\label{def:boundary}
 In a closure space $(X,\closure)$, the \emph{boundary} of $A \subseteq X$ is defined as $\boundary(A) = \closure(A) \setminus \interior(A)$. The \emph{interior boundary} is $\iboundary(A) = A \setminus \interior(A)$, and the \emph{closure boundary} is $\cboundary(A) = \closure(A) \setminus A$.
\end{definition}

\begin{proposition}\label{pro:boundary-properties}
 The following equations hold in a closure space:
 \begin{align}
    \boundary(A) & =  \cboundary(A) \cup \iboundary(A)  \label{eqn:boundary-properties-1}\\
    \cboundary(A) \cap \iboundary(A) & = \emptyset \label{eqn:boundary-properties-2} \\
    \boundary(A) & =  \boundary(\overline A) \label{eqn:boundary-properties-3} \\
    \cboundary(A) & =  \iboundary(\overline A)  \label{eqn:boundary-properties-4} \\
    \cboundary(A) & =  \boundary(A) \cap \overline A \label{eqn:boundary-properties-5} \\
    \iboundary(A) & =  \boundary(A) \cap A \label{eqn:boundary-properties-6} \\
    \boundary(A) & =  \closure(A) \cap \closure(\overline A) \label{eqn:boundary-properties-7}
 \end{align}
\end{proposition}

%% file: quasi-discrete.tex
\section{Quasi-discrete closure spaces}

In this section we see how a closure space may be derived starting from a \emph{binary relation}, that is, a \emph{graph}. The following comes from \cite{Gal03}.

\begin{definition}\label{def:closure-operator-of-a-relation}
Consider a set $X$ and a relation $R \subseteq X \times X$. A closure operator is obtained from $R$ as
$\closure_R(A) = A \cup \{ x \in X \mid \exists a \in A . (a,x) \in R \}$.
\end{definition}


\begin{remark}
 One could also change \autoref{def:closure-operator-of-a-relation} so that $\closure_R(A) = A \cup \{ x \in X \mid \exists a \in A . (x,a) \in R \}$, which actually is the definition of \cite{Gal03}. This does not affect the theory presented in the paper. Indeed, one obtains the same results by replacing $R$ with $R^{-1}$ in statements of theorems that explicitly use $R$, and are not invariant under such change. By our choice, closure represents the ``least possible enlargement'' of a set of nodes.
\end{remark}

\begin{proposition}\label{pro:closure-space-of-a-relation}
 The pair $(X,\closure_R)$ is a closure space.
\end{proposition}

Closure operators obtained by \autoref{def:closure-operator-of-a-relation} are not necessarily idempotent. Lemma 11 in \cite{Gal03} provides a necessary and sufficient condition, that we rephrase below. We let $R^=$ denote the reflexive closure of $R$ (that is, the least relation that includes $R$ and is reflexive).

\begin{lemma}\label{lem:idempotency-and-transitivity}
 $\closure_R$ is idempotent if and only if $R^=$ is transitive.
\end{lemma}

Note that, when $R$ is transitive, so is $R^=$, thus $\closure_R$ is idempotent. The vice-versa is not true, e.g., when $(x,y) \in R$, $(y,x) \in R$, but $(x,x) \notin R$. 

\begin{remark}\label{rem:open-sets-coarse}
In topology, open sets play a fundamental role. However, the situation is different in closure spaces derived from a relation $R$. For example, in the case of a closure space derived from a connected symmetric relation, the only open sets are the whole space, and the empty set.
\end{remark}
%

\begin{proposition}\label{pro:interior-boundary-in-quasi-discrete}
 Given $R \subseteq X \times X$, in the space $(X,\closure_R)$, we have:
 \begin{align}
 \interior(A) & = \{ x \in A \mid \lnot \exists a \in \overline A . (a,x) \in R \} \label{eqn:boundary-quasi-discrete-1} \\
 \iboundary(A) & = \{ x \in A \mid \exists a \in \overline A . (a,x) \in R\} \label{eqn:boundary-quasi-discrete-2} \\
 \cboundary(A) & = \{ x \in \overline A \mid \exists a \in A . (a,x) \in R \} \label{eqn:boundary-quasi-discrete-3}
 \end{align}
\end{proposition}

We note in passing that \cite{Gal99} provides an alternative definition of boundaries for closure spaces obtained from \autoref{def:closure-operator-of-a-relation}, and proves that it coincides with the topological definition (our \autoref{def:boundary}).
Closure spaces derived from a relation can be characterised as \emph{quasi-discrete} spaces (see also Lemma 9 of \cite{Gal03} and the subsequent statements).

\begin{definition}\label{def:quasi-discrete-closure-space}
 A closure space is \emph{quasi-discrete} if and only if one of the following equivalent conditions holds: i) each $x \in X$ has a \emph{minimal neighbourhood}\footnote{A \emph{minimal neighbourhood} of $x$ is a set that is a neighbourhood of $x$ (\autoref{def:closure-concepts}~(\ref{def:closure-concepts:neighbourhood})) and is included in all other neighbourhoods of $x$.} $\mneigh_x$; ii) for each $A \subseteq X$, $\closure(A) = \bigcup_{a \in A} \closure(\{a\})$.
\end{definition}

\noindent The following is shown as Theorem 1 in \cite{Gal03}.

\begin{theorem}
 A closure space $(X,\closure)$ is quasi-discrete if and only if there is a relation $R \subseteq X \times X$ such that $\closure = \closure_R$.
\end{theorem}


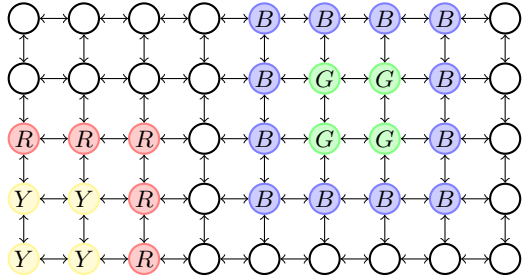
\begin{figure}[tbp]
\begin{center}
\begin{tikzpicture}{}

\foreach \i /\j / \fc / \bc/\c in { 
	1/1/yellow!50/yellow!20/Y, 
	2/1/yellow!50/yellow!20/Y, 
	3/1/red!50/red!20/R,
	4/1/black/white/\ ,
	5/1/black/white/\ ,
	6/1/black/white/\ ,
	7/1/black/white/\ ,
	8/1/black/white/\ ,
	9/1/black/white/\ ,
	1/2/yellow!50/yellow!20/Y, 
	2/2/yellow!50/yellow!20/Y, 
	3/2/red!50/red!20/R,
	4/2/black/white/\ ,
	5/2/blue!50/blue!20/B,
	6/2/blue!50/blue!20/B,
	7/2/blue!50/blue!20/B,
	8/2/blue!50/blue!20/B,
	9/2/black/white/\ ,
	1/3/red!50/red!20/R, 
	2/3/red!50/red!20/R, 
	3/3/red!50/red!20/R,
	4/3/black/white/\ ,
	5/3/blue!50/blue!20/B,
	6/3/green!50/green!20/G,
	7/3/green!50/green!20/G,
	8/3/blue!50/blue!20/B,
	9/3/black/white/\ ,
	1/4/black/white/\ , 
	2/4/black/white/\ , 
	3/4/black/white/\ ,
	4/4/black/white/\ ,
	5/4/blue!50/blue!20/B,
	6/4/green!50/green!20/G,
	7/4/green!50/green!20/G,
	8/4/blue!50/blue!20/B,
	9/4/black/white/\ ,
	1/5/black/white/\ , 
	2/5/black/white/\ , 
	3/5/black/white/\ ,
	4/5/black/white/\ ,
	5/5/blue!50/blue!20/B,
	6/5/blue!50/blue!20/B,
	7/5/blue!50/blue!20/B,
	8/5/blue!50/blue!20/B,
	9/5/black/white/\ 
	} {
		\node [circle,draw=\fc,fill=\bc,thick,inner sep=0pt,minimum size=4mm] at (\i*0.8,\j*0.8) (node\i\j) {$\c$};
}

\foreach \i [evaluate = \j as \ipp using int(\i+1)] in {1,2,3,4,5,6,7,8,9} {
	\foreach \j [evaluate = \j as \jpp using int(\j+1)] in {1,2,3,4,5} {
		\ifnum\j<5
		\draw[<->] (node\i\j) -- (node\i\jpp);
		\fi
		\ifnum\i<9
		\draw[<->] (node\i\j) -- (node\ipp\j);
		\fi
	}
}
\end{tikzpicture}
\end{center}
\caption{\label{fig:qdspace1}A  graph inducing a \emph{quasi-discrete} closure space}
\end{figure}


\begin{example}{}
\label{ex:qdspace1}
Every graph induces a \emph{quasi-discrete} closure space. For instance, we can consider the 
(undirected) graph depicted in \autoref{fig:qdspace1}. Let $R$ be the (symmetric) binary relation 
induced by the graph edges, and let $Y$ and $G$ denote the set 
of \emph{yellow} and \emph{green} nodes, respectively.
The closure $\closure_{R}(Y)$ consists of all \emph{yellow} and \emph{red} nodes, while the closure
$\closure_{R}(G)$ contains all \emph{green} and \emph{blue} nodes.
The interior $\interior(Y)$ of  $Y$ contains a single node, i.e. the one located at the 
bottom-left in \autoref{fig:qdspace1}. On the contrary, the \emph{interior} $\interior(G)$ of $G$ is empty.
Indeed, we have that $\boundary(G)=\closure(G)$, while $\iboundary(G)=G$ and $\cboundary(G)$ consists
of the \emph{blue} nodes.
\end{example}

%% file: digital-spatial-logic.tex
\section{A Spatial Logic for Closure Spaces}
\label{sec:dsl}

In this section we present a spatial logic that can be used to express properties of 
closure spaces. The logic features two \emph{spatial operators}:
a ``one step'' modality, turning closure into a logical operator, and a binary \emph{until} operator, which is interpreted spatially. 
Before introducing the complete framework, we first discuss the design of an \emph{until operator} $\phi \until \psi$.

The spatial logical operator $\until$ is interpreted on points of a closure space. The basic idea is that point $x$ satisfies $\phi \until \psi$ whenever it is included in an area $A$ satisfying $\phi$, and there is ``no way out'' from $A$ unless passing through an area $B$ that satisfies $\psi$.
%
For instance, if we consider the model of \autoref{fig:qdspace1}, \emph{yellow} nodes satisfy $yellow~\until~red$
while \emph{green} nodes satisfy $green~\until~blue$. 
%
%
To turn this intuition into a mathematical definition, one should clarify the meaning of the words \emph{area}, \emph{included}, \emph{passing}, in the context of closure spaces. 

In order to formally define our logic, and the \emph{until} operator in particular,  we first need to introduce 
the notion of \emph{model}, providing a context of evaluation for the satisfaction relation, as in 
$\model, x \models \phi \until \psi$.
From now on, fix a (finite or countable) set $\props$ of \emph{proposition letters.}

\begin{definition}\label{def:model}
A \emph{closure model} is a pair $\model = ((X,\closure),\eval)$ consisting of a closure space $(X,\closure)$ and a valuation $\eval : \props \to 2^X$, assigning to each  proposition letter the set of points where the proposition holds.
\end{definition}
 
 When $(X,\closure)$ is a topological space (that is, $\closure$ is idempotent), we call $\model$ a \emph{topological model}, in line with  \cite{vBB07}, and \cite{A02}, where the \emph{topological until} operator is presented. We recall it below.

\begin{definition}\label{def:until-t}
 The \emph{topological until} operator $\until_T$ is interpreted in a topological model $\model$ as $\model,x \models \phi \until_T \psi \iff \exists A$ open $ . x \in A \land \forall y \in A . \model, y \models \phi \land \forall z \in \boundary(A) . \model, z \models \psi$. 
\end{definition}

The intuition behind this definition is that one seeks for an area $A$ (which, topologically speaking, could sensibly be an open set) where $\phi$ holds, and that is completely surrounded by points where $\psi$ holds. 
Unfortunately, \autoref{def:until-t} cannot be translated directly to closure spaces, even if all the used topological notions have a counterpart in the more general setting of closure spaces. Open sets in closure spaces are often too coarse (see \autoref{rem:open-sets-coarse}). 
For this reason, we can modify \autoref{def:until-t} by not requiring $A$ to be an \emph{open} set.
However, the usage of $\boundary$ in \autoref{def:until-t} is not satisfactory either. By \autoref{pro:boundary-properties} we have $\boundary(A) = \cboundary(A) \cup \iboundary(A)$, where 
$\iboundary(A)$ is included in $A$ while $\cboundary(A)$ is in $\overline{A}$. 
For instance, when $\boundary$ is used in \autoref{def:until-t}, we have that the \emph{green} nodes 
in \autoref{fig:qdspace1} do not satisfy $green~\until_{T}~blue$. Indeed, as we remarked in \autoref{ex:qdspace1}, 
the \emph{boundary} of the set $G$ of green nodes coincide with the closure of $G$ that contains both \emph{green}
and \emph{blue} nodes.

A more satisfactory definition can be obtained by letting $\cboundary$ play the same role as $\boundary$ in \autoref{def:until-t} and not requiring $A$ to be an open set. We shall in fact require that $\phi$ is satisfied by \emph{all} the points of $A$, and that in $\cboundary(A)$, $\psi$ holds. This allows us to ensure that there are no ``gaps'' between the
region satisfying $\phi$ and that satisfying $\psi$. 

\subsection{Syntax and Semantics of $\slcs$}

We can now define $\slcs$: a \emph{Spatial Logic for Closure Spaces}. The logic features boolean operators, a ``one step'' modality, turning closure into a logical operator, and a spatially interpreted \emph{until} operator. More precisely, as we shall see, the $\slcs$ formula $\phi \until \psi$ requires $\phi$ to hold at least on one point. The operator is similar to a \emph{weak until} in temporal logics terminology, as there may be no point satisfying $\psi$, if $\phi$ holds everywhere. 

\begin{definition}\label{def:logic-syntax}
 The syntax of $\slcs$ is defined by the following grammar, where $p$ ranges over $\props$:
 $$ \Phi ::= p \mid \top \mid \lnot \Phi \mid \Phi \land \Phi \mid \lozenge \Phi \mid \Phi \until \Phi $$
\end{definition}

Here, $\top$ denotes \emph{true}, $\lnot$ is negation, $\land$ is conjunction, $\lozenge$ is the \emph{closure} operator, and $\until$ is the \emph{until} operator. Closure (and interior, see \autoref{fig:derivableoperators}) operators come from the tradition of topological spatial logics \cite{vBB07}. 

\begin{definition}\label{def:closure-semantics}
 Satisfaction $\model,x \models \phi$ of formula $\phi$ at point $x$ in model $\model = ((X,\closure),\eval)$ is defined, by induction on terms, as follows:
 \[\begin{array}{rclcl}
   \model,x & \models & p & \iff & x \in \eval(p) \\
   \model,x & \models & \top & \iff & \mathit{true} \\
   \model,x & \models & \lnot \phi & \iff & \model,x \not \models \phi \\
   \model,x & \models & \phi \land \psi & \iff & \model,x \models \phi \text{ and } \model, x \models \psi \\
   \model,x & \models & \lozenge \phi & \iff & x \in \closure(\{ y \in X | \model, y \models \phi \}) \\
   \model,x & \models & \phi \until \psi & \iff & \exists A \subseteq X . x \in A \land \forall y \in A . \model, y \models \phi \land \\ & & & & \qquad \land \forall z \in \cboundary(A) . \model, z \models \psi
 \end{array}\]
\end{definition}

\begin{figure}[tbp]
 \[\begin{array}{lclclcl}
    \bot & \triangleq & \lnot \top & \qquad \qquad &
    \phi \lor \psi & \triangleq & \lnot (\lnot \phi \land \lnot \psi) 
    \\
    \square \phi & \triangleq & \lnot (\lozenge \lnot \phi) & 
    &
    \lboundary \phi & \triangleq & (\lozenge \phi) \land (\lnot \square \phi) 
    \\
    \liboundary \phi & \triangleq & \phi \land (\lnot \square \phi) & 
    &
    \lcboundary \phi & \triangleq & (\lozenge \phi) \land (\lnot \phi) 
    \\
    \phi \ldualuntil \psi & \triangleq & \neg ( (\neg \psi) \until (\neg \phi) ) & 
    & \leverywhere \phi & \triangleq & \phi \until \bot 
    \\  
    \lsomewhere \phi & \triangleq & \neg \leverywhere (\neg \phi) & 
 \end{array}\]
\caption{\label{fig:derivableoperators}$\slcs$ derivable operators}
\end{figure}

In \autoref{fig:derivableoperators}, we present some derived operators. Besides standard logical connectives, the logic can express the \emph{interior} ($\square\phi$), the \emph{boundary} ($\lboundary \phi$),
the \emph{interior boundary} ($\liboundary \phi$) and the \emph{closure boundary} ($\lcboundary \phi$) 
of the set of points satisfying formula $\phi$. Moreover, by appropriately using the \emph{until} operator, operators concerning
\emph{reachability} ($\phi \ldualuntil \psi$), \emph{global satisfaction} ($\leverywhere \phi$) and \emph{possible satisfaction} 
($\lsomewhere \phi$) can be derived.

To clarify the expressive power of $\until$ and operators derived from it we provide
\autoref{thm:simple-discrete-paths-until} and \autoref{thm:other-side-of-simple-discrete-paths-until}, giving a formal meaning to the idea of ``way out'' of $\phi$, 
and providing an interpretation of $\until$ in terms of \emph{paths}.

\begin{definition}\label{def:continuous-function}
 A \emph{closure-continuous function} $f : (X_1,\closure_1) \to (X_2,\closure_2)$ is a function $f : X_1 \to X_2$ such that, for all $A \subseteq X_1$, $f(\closure_1(A)) \subseteq \closure_2(f(A))$.
\end{definition}

\begin{definition}\label{def:path}
 Consider a closure space $(X,\closure)$, and the quasi-discrete space $(\nats,\closure_\succ)$, where $(n,m) \in \succ \iff m = n+1$. A (countable) \emph{path} in $(X,\closure)$ is a closure-continuous function $p : (\nats,\closure_\succ) \to (X,\closure)$. 
 We call $p$ a path \emph{from} $x$, and write $p : x \pto{}{}$, when $p(0) = x$. We write $y \in p$ whenever there is $l \in \nats$ such that $p(l) = y$. We write $p : x \pto{A}{y}$ when $p$ is a path from $x$, and there is $l$ with $p(l) = y$ and for all $l' \leq l . p(l') \in A$.
\end{definition}

\begin{theorem}\label{thm:simple-discrete-paths-until}
 If $\model,x \models \phi \until \psi$, then for each $p : x \pto {}{}$ and $l$, if $\model, p(l) \models \lnot \phi$, there is $k \in \{1, \ldots, l\}$ such that $\model,p(k) \models \psi$. 
\end{theorem}

\noindent \autoref{thm:simple-discrete-paths-until} can be strengthened to a necessary and sufficient condition in the case of models based on quasi-discrete spaces. First, we establish that paths in a quasi-discrete space are also paths in its underlying graph.

\begin{lemma}\label{lem:paths-are-paths}
 Given path $p$ in a quasi-discrete space $(X, \closure_R)$, for all $i \in \nats$ with $p(i) \neq p(i+1)$, we have $(p(i),p(i+1)) \in R$, i.e., the image of $p$ is a (graph theoretical, infinite) path in the graph of $R$. Conversely, each path in the graph of $R$ uniquely determines a path in the sense of \autoref{def:path}.
\end{lemma}

\begin{theorem}\label{thm:other-side-of-simple-discrete-paths-until}
 In a quasi-discrete closure model $\model$, $\model,x \models \phi \until \psi$  if and only if $\model, x \models \phi$, and for each path $p : x \pto {}{}$ and $l\in \nats$, if $\model, p(l) \models \lnot \phi$, there is $k \in \{1, \ldots, l\}$ such that $\model,p(k) \models \psi$. 
\end{theorem}

\begin{remark}{}
\label{remark:duals}
Directly from \autoref{thm:other-side-of-simple-discrete-paths-until} and from the definitions in \autoref{fig:derivableoperators} we have also that in a quasi-discrete closure model $\model$:
\begin{enumerate}
\item $\model,x \models \phi \ldualuntil \psi$  iff. there is $p : x \pto {}{}$ and $k\in \nats$ such that $\model, p(k) \models \psi$ and for each $j\in \{1,\ldots,k\}$ $\model, p(j) \models \phi$;
\item $\model,x \models \leverywhere \phi$  iff. for each $p : x \pto {}{}$ and $i\in \nats$, $\model, p(i) \models \phi$; 
\item $\model,x \models \lsomewhere \phi$  iff. there is $p : x \pto {}{}$ and $i\in \nats$ such that $\model, p(i) \models \phi$.
\end{enumerate} 
\end{remark}

Note that, 
a point $x$ 
satisfies $\phi \ldualuntil \psi$ if and only if either $\psi$ is satisfied by $x$ or there exists a sequence of
points after $x$, all satisfying $\phi$, leading to a point satisfying both $\psi$ and $\phi$. In the second case,
it is not required that $x$ satisfies $\phi$.

%% file: model-checking.tex
\section{Model checking $\slcs$ formulas}\label{sec:model-checking}

In this section we present a model checking algorithm for $\slcs$, which is
a variant of standard CTL 
model
checking~\cite{BK08}.
%
\begin{algorithm}[tbp]
\myfun{$\check(\model,\phi)$}{
  
  \KwIn{Quasi-discrete closure model $\model = ((X,\closure),\eval)$, $\slcs$ formula $\phi$}
  \KwOut{Set of points $\{ x \in X \mid \model, x \models \phi\}$}
  
  \Match{$\phi$}
  {
     \lCase{$\top:$}{\Return $X$}
     \lCase{$p:$}{\Return $\eval(p)$}
     \Case{$\lnot \psi:$}{
	\Let $P = \check(\model,\psi)$ \In\; 
	\Return $X \setminus P$
      }
     \Case{$\psi \land \xi:$}{
	\Let $P = \check(\model,\psi)$ \In\;
	\Let $Q = \check(\model,\xi)$ \In\;
	\Return $P \cap Q$
      }
     \Case{$\lozenge \psi:$}{
	\Let $P = \check(\model,\psi)$ \In\;
	\Return $\closure(P)$
      }

     \lCase{$\psi \until \xi:$}{
        \Return \checkUntil($\model$,$\psi$, $\xi$)
     \vskip 4pt
     }
   }   
  }
%
%
%
\caption{\label{alg:decision-procedure}Decision procedure for the model checking problem.}
\end{algorithm}
Function \check, presented in \autoref{alg:decision-procedure}, takes as input a finite quasi-discrete model $\model=((X,\closure_{R}),\eval)$ and an $\slcs$ formula $\phi$, 
and returns the set of all points  in $X$ satisfying $\phi$. 
The function is inductively defined on the structure of $\phi$ and, 
following a bottom-up approach, computes the resulting set via an
appropriate combination of the recursive invocations of \check on the sub-formulas  of $\phi$.
When $\phi$ is $\top$, $p$, $\lnot \psi$ or $\psi\land \xi$, definition of $\check(\model,\phi)$ is 
as expected. 
To compute the set of points satisfying $\lozenge \psi$, the closure
operator $\closure$ of the space is applied to the set of points satisfying $\psi$.
%

When $\phi$ is of the form $\psi \until \xi$, function \check relies on
the function \checkUntil defined in \autoref{alg:check-until-quasi-discrete}.
This function takes as parameters a model $\model$ and two $\slcs$ formulas $\psi$ and $\xi$ and 
computes the set of points in $\model$ satisfying $\psi \until \xi$ by removing from $V=\check(\model,\psi)$
all the \emph{bad} points. A point is \emph{bad} if 
there exists a path passing through it, that leads to
a point satisfying $\neg\psi$ 
without passing through a point satisfying $\xi$.
Let $Q=\check(\model,\xi)$ be the set of points in $\model$ satisfying $\xi$. 
To identify the \emph{bad} points in $V$ the function \checkUntil performs a \emph{backward search} from 
$T=\cboundary(V\cup Q)$.  Note that any \emph{path} exiting from $V\cup Q$ has to pass through points
in $T$. Moreover, the latter only contains points that satisfy neither $\psi$ nor $\xi$.
Until $T$ is empty, function \checkUntil first picks an element $x$ in $T$ and then removes from $V$ 
the set of (bad) points $N$ that can reach  $x$ in \emph{one step}.  
To compute the set $N$ we use the function $pre(x) = \{ y \in X \mid  (y,x) \in R \}$.\footnote{Function $pre$
can be \emph{pre-computed} when the relation $R$ is loaded from the input.}
At the end of each iteration the set $T$ is updated by considering the set of new 
discovered \emph{bad points}.

%
%

%
\begin{lemma}
\label{lemma:termination}
Let $X$ a finite set and $R\subseteq X\times X$. 
 For any finite quasi-discrete model $\model=((X,\closure_{R}),\eval)$ and $\slcs$ formula $\phi$ 
with $k$ operators, 
 \check terminates  in $\mathcal{O}(k\cdot(|X|+|R|))$ steps.
\end{lemma}


%

\begin{algorithm}[btp]
        \myfun{\checkUntil($\model$,$\psi$, $\xi$)} {
	
	\Let $V = \check(\model,\psi)$ \In\;
	\Let $Q = \check(\model,\xi)$ \In\;
 	\Var $T$ := $\cboundary(V \cup Q)$ \;
  
	\While{$T \neq \emptyset$}{
        $T'$ := $\emptyset$\;
		\For{$x\in T$}{
              $N$ := $pre(x) \cap V$\; 
              $V$ := $V \setminus N$\;
              $T'$ := $T'\cup (N \setminus Q)$\;
		}                
        $T$ := $T'$;
        }
  	\Return $V$

     }
~ \\[20pt] 
\caption{\label{alg:check-until-quasi-discrete}Checking until formulas in a quasi-discrete closure space.}
\end{algorithm}

%
\begin{theorem}\label{thm:sound-compl}
For any finite quasi-discrete closure model $\model=((X,\closure),\eval)$ and $\slcs$ formula $\phi$, $x\in \check(\model,\phi)$
if and only if $\model,x\models \phi$.
\end{theorem}

%% file: examples.tex
\section{A model checker for spatial logics}\label{sec:discrete-examples}

The algorithm described in \autoref{sec:model-checking} is available as a proof-of-concept tool\footnote{Web site: \url{http://www.github.com/vincenzoml/slcs}.}. The tool, implemented using the functional language OCaml, contains a generic implementation of a global model-checker using closure spaces, parametrised by  the type of models.

An example of the tool usage is to approximately identify regions of interest on a digital picture (e.g., a map, or a medical image), using spatial formulas. In this case, digital pictures are treated as quasi-discrete models in the plane $\ints \times \ints$. The language of propositions is extended to simple formulas dealing with colour ranges, in order to cope with images where there are different shades of certain colours.

\begin{figure}[tbp]
\centering{
 \begin{minipage}[t]{.35\textwidth}
    \includegraphics[width=\textwidth]{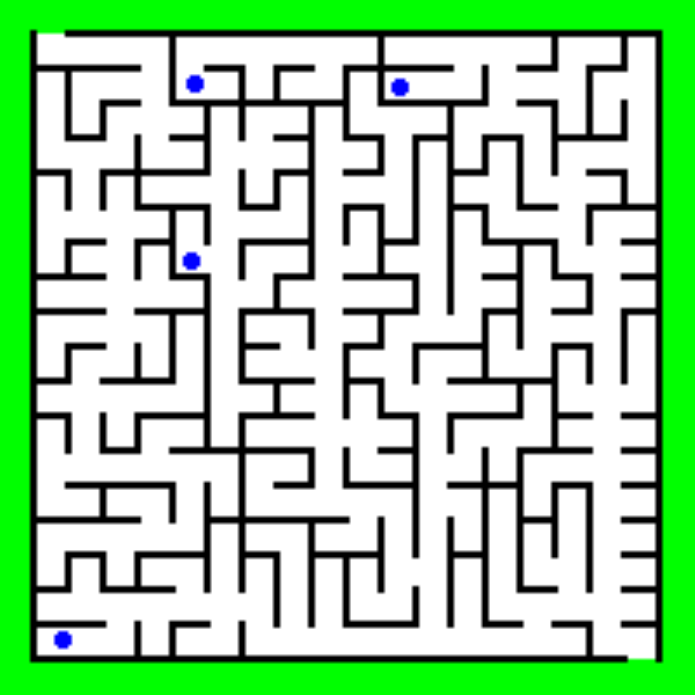}
    \caption{\label{fig:maze} A maze. }
 \end{minipage} \hskip 10pt
 \begin{minipage}[t]{.35\textwidth}
    \includegraphics[width=\textwidth]{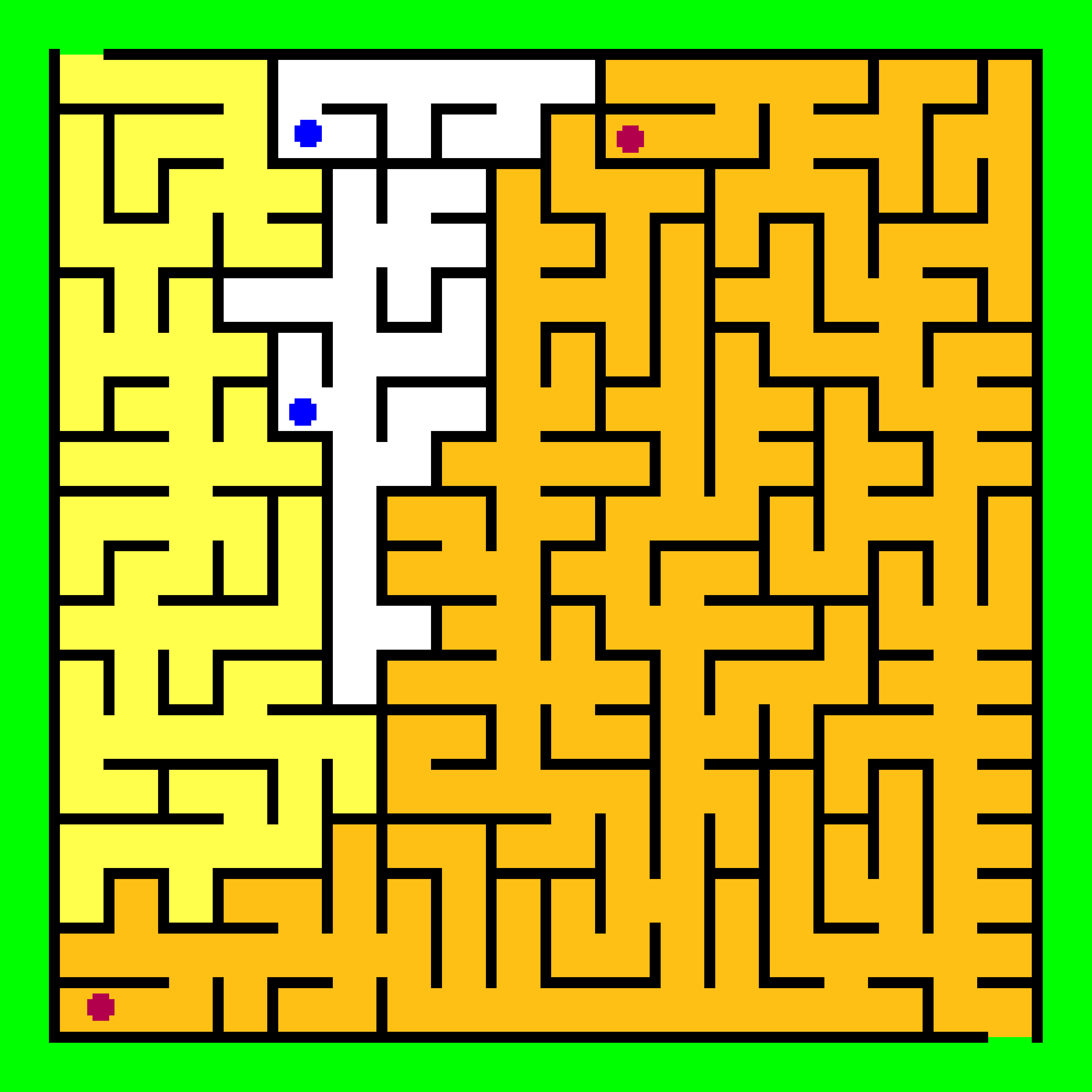}
    \caption{\label{fig:maze-out} Model checker output. }
 \end{minipage}
}
\end{figure}

 In \autoref{fig:maze} we show a digital picture of a maze. The green area is the exit. The blue areas are start points. The input of the tool is shown in \autoref{fig:maze-code}, where the \verb.Paint. command is used to invoke the global model checker and colour points satisfying a given formula. Three formulas, making use of the until operator, are used to identify interesting areas. The output of the tool is in \autoref{fig:maze-out}. The colour red denotes start points from which the exit can be reached. Orange and yellow indicate the two regions through which the exit can be reached, including and excluding a start point, respectively.

\begin{figure}[!t]
\begin{minipage}{\textwidth}
\begin{verbatim}
Let reach(a,b) = !( (!b) U (!a) );
Let reachThrough(a,b) = a & reach((a|b),b);

Let toExit = reachThrough(["white"],["green"]);
Let fromStartToExit = toExit & reachThrough(["white"],["blue"]);
Let startCanExit = reachThrough(["blue"],fromStartToExit);

Paint "yellow" toExit;
Paint "orange" fromStartToExit;
Paint "red" startCanExit;
\end{verbatim}
  \vskip -10pt
  \caption{\label{fig:maze-code} Input to the model checker.}
  \vskip -10pt
\end{minipage}
\end{figure}

\begin{figure}[tbp]
\begin{center}
\begin{minipage}{.40\textwidth}
 \includegraphics[width=\textwidth]{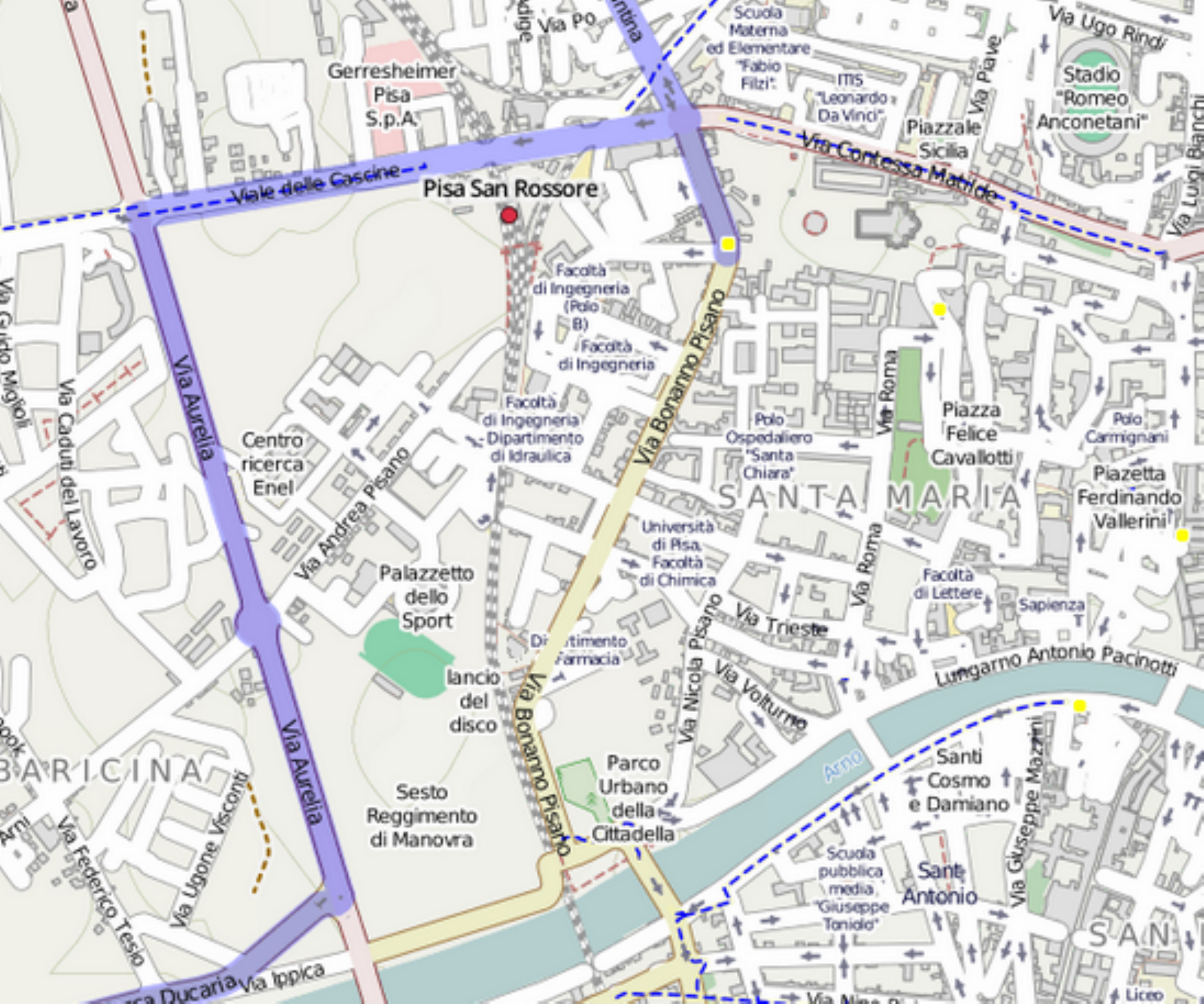}
 \caption{\label{fig:openstreetmap-pisa} Input: the map of a town.}
\end{minipage}
\begin{minipage}{.40\textwidth}
 \includegraphics[width=\textwidth]{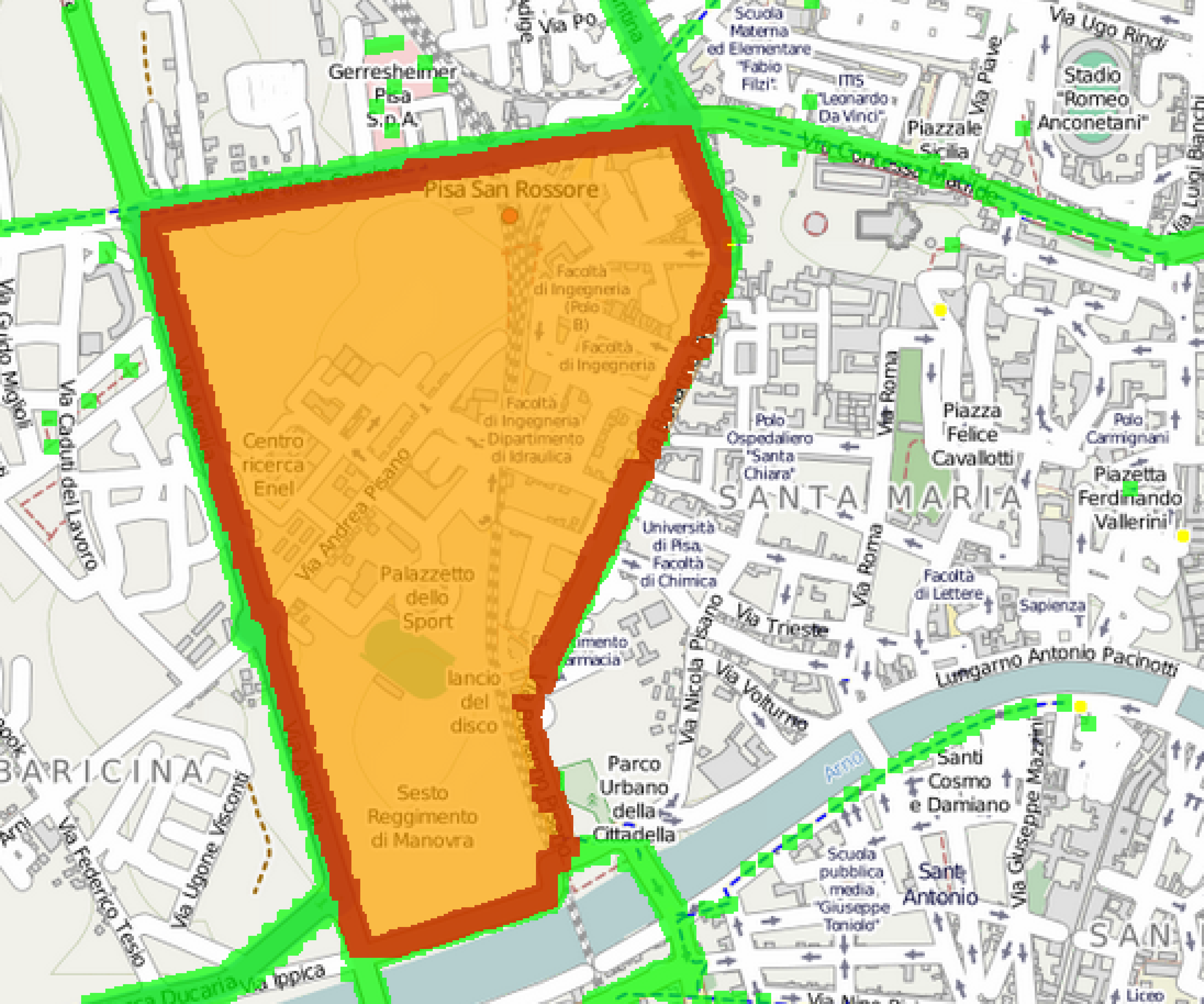}
 \caption{\label{fig:openstreetmap-pisa-out} Output of the tool.}
\end{minipage}
\end{center}
\end{figure}

 In \autoref{fig:openstreetmap-pisa} we show a digital image\footnote{\copyright \emph{OpenStreetMap contributors} -- \url{http://www.openstreetmap.org/copyright}.} depicting a portion of the map of Pisa, featuring a red circle which denotes a train station. Streets of different importance are painted with different colors in the map. The model checker is used to identify the area surrounding the station which is delimited by main streets, and the delimiting main streets. 
 The output of the tool is shown in \autoref{fig:openstreetmap-pisa-out}, where the station area is coloured in orange, the surrounding main streets are red, and other main streets are in green. 
We omit the source code of the model checking session for space reasons (see the source code of the tool).
%
As a mere hint on how practical it is to use a model checker for image analysis, the execution time on our test image, consisting of about 250000 pixels, is in the order of ten seconds on a standard laptop equipped with a 2Ghz processor.

%% file: future-work.tex
\section{Conclusions and Future Work}\label{sec:discussion}

In this paper, we have presented a methodology to verify properties that depend upon space. 
We have defined an appropriate logic, stemming from the tradition of topological interpretations of 
modal logics, dating back to earlier logicians such as Tarski, where modalities describe neighbourhood. 
The topological definitions have been lifted to a more general setting, also encompassing discrete, 
graph-based structures. 
The proposed framework has been extended with a spatial variant of the \emph{until} operator, 
and we have also defined an efficient model checking procedure, which is implemented in a proof-of-concept tool. 

As future work, we first of all plan to merge the results presented in this paper with temporal reasoning. 
This integration can be done in more than one way. It is not difficult to consider ``snapshot'' models consisting of a temporal model (e.g., a Kripke frame) where each state is in turn a closure model, and atomic formulas of the temporal fragment are replaced by spatial formulas. The various possible combinations of temporal and spatial operators, in linear and branching time, are examined for the case of topological models and basic modal formulas in \cite{KKWZ07}. 
Snapshot models may be susceptible to state-space explosion problems as spatial formulas could need to
be recomputed at every state. On the other hand, one might be able to exploit the fact that changes of space
over time are incremental and local in nature.
Promising ideas are presented both in \cite{Gal03}, where principles of ``continuous change'' are proposed in the setting of closure spaces, and in \cite{KM07} where spatio-temporal models are generated by locally-scoped update functions, in order to describe dynamic systems. 
In the setting of collective adaptive systems, it will be certainly needed to extend the basic framework we presented with metric aspects (e.g., distance-bounded variants of the until operator), and probabilistic aspects, using atomic formulas that are probability distributions. A thorough investigation of these issues will be the object of future research.

A challenge in spatial and spatio-temporal reasoning is posed by recursive spatial formulas, \emph{a la} $\mu$-calculus, especially on infinite structures with relatively straightforward generating functions (think of fractals, or fluid flow analysis of continuous structures). Such infinite structures could be described by topologically enhanced variants of $\omega$-automata. Classes of automata exist living in specific topological structures; an example is given by \emph{nominal automata} (see e.g., \cite{BKL11,GC11,KST12}), that can be defined using presheaf toposes \cite{fs06}.
 This standpoint could be enhanced with notions of neighbourhood coming from closure spaces, with the aim of developing a unifying theory of languages and automata describing space, graphs, and process calculi with resources.

%% file: proofs.tex
\section{Proofs}

\begin{proof}(of \autoref{lem:interior-monotone})
%

Proof of \autoref{lem:interior-monotone-1}
 $
  \der A = \interior(A)
  \stp \iff \overline A = \overline{\interior(A)} 
  \stp \iff \overline A = \closure(\overline A)
 $

Proof of \autoref{lem:interior-monotone-2}
 $ \der A \subseteq B
   \stp \iff A \cup B = B
   \jstp \implies {$def. closure$} \closure(A) \cup \closure(B) = \closure(B)
   \stp \iff \closure(A) \subseteq \closure(B)
 $
 
 $ \der A \subseteq B
   \stp \implies \overline B \subseteq \overline A
   \jstp \implies {$previous part of the proof$} \closure(\overline B) \subseteq \closure(\overline A) 
   \stp \iff \overline{\interior(B)} \subseteq \overline{\interior(A)}
   \stp \iff \interior(A) \subseteq \interior(B)
 $

Proof of \autoref{lem:interior-monotone-3}
 $\der \interior ( A \cap B ) 
  \stp = \overline{\closure(\overline{A \cap B})} 
  \stp = \overline{\closure(\overline A \cup \overline B)}
  \jstp = {$definition of closure$} \overline{\closure(\overline A) \cup \closure(\overline B)}
  \stp = \overline{\closure(\overline A)} \cap \overline{\closure(\overline B)}
  \stp = \interior(A) \cap \interior(B) 
  \jstp = {A$ and $B$ are open$} A \cap B
 $
 
 Finally, we prove that, whenever all sets in a collection $A_{i \in I}$ are open, we have $\interior(\bigcup_{i \in I} A_i) = \bigcup_{i \in I} A_i$, that is, the union of open sets is open. The left-to-right inclusion is true since $\forall A . \interior(A) \subseteq A$, which is the property $\forall A . A \subseteq \closure(A)$ (\autoref{def:closure-space}), dualised by the definition of interior. For the right-to-left inclusion we have:
%
 $\der true
  \jstp \implies {$definition of $ \bigcup} \forall i \in I . A_i \subseteq \bigcup_{i \in I} A_i
  \jstp \implies {\interior$ is monotone by \autoref{lem:interior-monotone}, \autoref{lem:interior-monotone-2}$} 
      \forall i \in I . \interior(A_i) \subseteq \interior(\bigcup_{i \in I} A_i)
  \jstp \implies {\forall i \in I . A_i$ is open$} \forall i \in I . A_i \subseteq \interior(\bigcup_{i \in I} A_i)
  \stp \implies \bigcup_{i \in I} A_i \subseteq \interior(\bigcup_{i \in I} A_i)
 $

\end{proof}

%
%

\begin{proof}(of \autoref{pro:boundary-properties})

\autoref{eqn:boundary-properties-1}:
$
  \der \boundary(A)
  \stp = \closure(A) \setminus \interior(A)
  \jstp = {\interior(A) \subseteq A, \forall A,B,C . B \subseteq C \implies A \setminus B = (A \setminus C) \cup (C \setminus B)} (\closure(A) \setminus A) \cup (A \setminus \interior(A))
  \stp = \cboundary(A) \cup \iboundary(A)
 $ 

 \medskip
 
 \autoref{eqn:boundary-properties-2}:
 $
  \der \cboundary(A) \cap \iboundary(A) 
  \stp = (\closure(A) \setminus A) \cap (A \setminus \interior(A))
  \jstp = {\closure(A) \setminus A \subseteq \overline A, A \setminus \interior(A) \subseteq A} \emptyset
 $
 
 \medskip
 
 \autoref{eqn:boundary-properties-3}:
 $
  \der
  \boundary(A) 
  \stp = \closure(A) \setminus \interior(A) 
  \stp = \overline{\interior(\overline A)} \setminus \overline{\closure(\overline A)}
  \stp = \closure(\overline A) \setminus \interior(\overline A)
  \stp = \boundary(\overline A)
 $
 
 \medskip
 
 \autoref{eqn:boundary-properties-4}:
 $
   \der \iboundary(\overline{A})
   \stp = \overline A \setminus \interior(\overline A)
   \stp = \overline A \setminus \overline{\closure(A)}
   \stp = \closure(A) \setminus A
   \stp = \cboundary(A)
 $
 
 \medskip
 
 \autoref{eqn:boundary-properties-5}:
 $
  \der \cboundary(A) 
  \stp = \closure(A) \setminus A
  \jstp = {\interior(A) \subseteq A}
          (\closure(A) \setminus \interior(A)) \setminus A
  \stp = \boundary(A) \setminus A
  \stp = \boundary(A) \cap \overline A
 $
 
 \medskip
 
 \autoref{eqn:boundary-properties-6}:
 
 $
 \der \iboundary(A)
 \jstp = {$Statement \ref{eqn:boundary-properties-4}$}
         \cboundary(\overline A)
 \jstp = {$Statement \ref{eqn:boundary-properties-5}$}
         \boundary(\overline A) \cap A
 \jstp = {$Statement \ref{eqn:boundary-properties-3}$}
         \boundary(A) \cap A
 $
 
 \medskip
 
 \autoref{eqn:boundary-properties-7}:
 $
 \der \boundary(A) 
 \stp = \closure(A) \setminus \interior(A) 
 \stp = \closure(A) \cap \overline{\interior(A)} 
 \stp = \closure(A) \cap \closure(\overline A)
 $
\end{proof}


\begin{proof}(of \autoref{pro:closure-space-of-a-relation})
\medskip

  Axiom \ref{def:closure-space:closure-of_emptyset}:
 $\der \closure_R(\emptyset) = \emptyset \cup \{ x \in X \mid \exists a \in \emptyset . (a,x) \in R\} = \emptyset$
 \medskip
 
 Axiom \ref{def:closure-space:closure-larger}:
 $\der A 
  \jstp \subseteq {A \subseteq A \cup B} \closure_R(A)$
 \medskip
 
 Axiom \ref{def:closure-space:closure-union}:
 $\der \closure_R(A \cup B) 
  \stp = A \cup B \cup \{ x \in X \mid \exists c \in A \cup B . (c,x) \in R \} 
  \jstp = { c \in A \cup B \iff c \in A \lor c \in B }
         A \cup B \cup \{ x \in X \mid \exists c \in A . (c,x) \in R \} \cup \{ x \in X \mid \exists c \in B . (c,x) \in R \} 
  \stp =  \closure_R(A) \cup \closure_R(B)
 $
\end{proof}

\begin{proof}(of \autoref{pro:interior-boundary-in-quasi-discrete})

 \autoref{eqn:boundary-quasi-discrete-1}:
 $
 \der \interior(A) 
 \stp = \overline{\closure_R(\overline A)}
 \stp = \overline{\overline A \cup \{x \in X \mid \exists a \in \overline A . (a,x) \in R \}}
 \stp = A \cap \{ x \in X \mid  \lnot \exists a \in \overline A . (a,x) \in R\} 
 \stp = \{ x \in A \mid \lnot \exists a \in \overline A . (a,x) \in R \}
 $
 
 \autoref{eqn:boundary-quasi-discrete-2}:
 $
 \der \iboundary(A) 
 \stp = A \setminus \interior(A) 
 \stp = A \setminus \{ x \in A \mid \lnot \exists a \in \overline A . (a,x) \in R \}
 \stp = A \cap \{ x \in A \mid \exists a \in \overline A . (a,x) \in R \}
 \stp = \{ x \in A \mid \exists a \in \overline A . (a,x) \in R \}
 $
 
 \autoref{eqn:boundary-quasi-discrete-3}:
 $
 \der \cboundary(A) 
 \stp = \closure(A) \setminus A 
 \stp = (A \cup \{ x \in X \mid \exists a \in A . (a,x) \in R \} ) \setminus A 
 \stp = (A \cup \{ x \in X \mid \exists a \in A . (a,x) \in R \}) \cap \overline A
 \stp = (A \cap \overline A) \cup (\{ x \in X \mid \exists a \in A . (a,x) \in R \} \cap \overline A) 
 \stp = \{ x \in \overline A \mid \exists a \in A . (a,x) \in R \}
 $
\end{proof}

\begin{proof}(of \autoref{thm:simple-discrete-paths-until})
 Let $\model = ((X,\closure),\eval)$. Since $\model, x \models \phi \until \psi$, let $A$ be the set from \autoref{def:closure-semantics}. Let $p : x \pto{}{}$, and $l$ be such that $\model, p(l) \models \lnot \phi$. Consider the set $K^- = \{ k \mid \forall h \in \{0,\ldots,k\} . p(h) \in A \}$.  Since $0 \in K^-$, we have $K^- \neq \emptyset$. Consider the complement of $K^-$, namely $K^+ = \nats \setminus K^-$.  Since all points in $A$ satisfy $\phi$, and $p(l) \models \lnot \phi$, we have $l \in K^+$, thus $K^+ \neq \emptyset$. By existence of $l$, $K^-$ is finite, thus, being non-empty, it has a greatest element. Being a non-empty subset of the natural numbers, $K^+$ has a least element. Let $k^- = \max K^-$ and $k^+ = \min K^+$. Noting that if $k \in K^-$ and $h \in [0,k)$, then $h \in K^-$, we have $k^- + 1 = k^+$, thus $(k^-,k^+) \in \succ$. 
 Let $S = \{ p(k) | k \in K^- \} \subseteq A$. By monotonicity of closure, we have $\closure(S) \subseteq \closure(A)$. By definition of $\closure_\succ$, we have $k^+ \in \closure_\succ(K^-)$, thus by closure-continuity $p(k^+) \in \closure(S)$ and therefore $p(k^+) \in \closure(A)$.  But it is also true that $p(k^+) \notin A$; if $p(k^+) \in A$, then we would have $k^+ \in K^-$, by definition of $K^-$. Thus, $p(k^+) \in \cboundary(A)$, therefore $p(k^+) \models \psi$. Note that in particular $k^+ \neq 0$ as $p(0) = x \in A$, and $k^+ \leq l$ as $l \in K^+$ and $k^+ = \min K^+$.
\end{proof}

\begin{proof}(of \autoref{lem:paths-are-paths})
 For one direction of the proof, assume $p$ is a closure-continuous function. Importing definitions from \autoref{def:path} and the statement of \autoref{lem:paths-are-paths}, we have 
 $\der
 (i,i+1) \in \succ 
 \stp \implies i + 1 \in \closure_\succ(\{i\})
 \jstp \implies{ p $ closure-continuous$} p(i+1) \in \closure_R(p(\{i\})) 
 \stp \iff p(i+1) \in \closure_R(\{p(i)\}) 
 \stp \iff p(i+1) \in \{ p(i) \} \cup \{ x \mid (p(i),x) \in R \}
 \stp \iff p(i+1) = p(i) \lor (p(i),p(i+1)) \in R
 $
 
 ~ \\
 \noindent For the other direction, given a path $x_i$ of length $l$ in $R$, define $p(i) = x_i$. Closure-continuity of $p$ is straightforward.
\end{proof}

\begin{proof}(of \autoref{thm:other-side-of-simple-discrete-paths-until})
 One direction is given by \autoref{thm:simple-discrete-paths-until}. For the other direction, assume $\model = ((X,\closure_R),\eval)$ where $\closure_R$ is the closure operator derived by a relation $R$. Consider point $x$ with $\model, x \models \phi$, and assume that for each $p : x \pto{}{}$ and $l$ such that $\model, p(l) \models \lnot \phi$ there is $k \in \{1, \ldots, l\}$ such that $\model, p(k) \models \psi$. 
Define the following set: 
$$A_x = \{x\} \cup \{ y \in X \mid \exists p : x \pto{}{} . \exists l > 0. p(l) = y \land \forall k \in \{1,\ldots,l\} . \model, p(k) \models \phi \land \lnot \psi\}$$

  
 We will use $A_x$ as a witness of the existence of a set $A$ in \autoref{def:closure-semantics}, in order to prove that $\model, x \models \phi \until \psi$. Note that by definition of $A_x$, $x \in A_x$ and $\forall y \in A_x . \model, p(y) \models \phi$. We need to show that $\forall z \in \cboundary(A_x) . \model, z \models \psi$. Consider $z \in \cboundary(A_x)$. Since $\model$ is based on a quasi-discrete closure space, by \autoref{eqn:boundary-quasi-discrete-3} in \autoref{pro:interior-boundary-in-quasi-discrete}, we have $z \in \overline{A_x}$ and there is $y \in A_x$ such that $(y,z) \in R$. Suppose $y = x$. Let $p$ be the path defined by $p(0) = x$, $p(i \neq 0) = z$. If $\model, z \models \phi$, suppose $\model, z \nmodels \psi$; then $z \in A_x$, witnessed by the path $p$, with $l=1$; therefore, since $z \in \overline {A_x}$ we have $\model, z \models \psi$. If $\model, z \nmodels \phi$, then noting $p(1) = z$, by hypothesis, there is $k \in \{1,\ldots,1\}$ with $\model, p(k) \models \psi$, that is $\model, z \models \psi$. Suppose $y \neq x$. Then there are $p : x \pto{}{}$  and $l > 0$ such that $p(l) = y \land \forall k \in \{1,\ldots,l\} . \model, p(k) \models \phi \land \lnot \psi$. Define $p'$ by $p'(l') = p(l')$ if $l' \leq l$, and $p'(l') = z$ otherwise. The rest of the proof mimics the case $y = x$. If $\model, z \models \phi$, then $\model, z \nmodels \psi$ implies $z \in A_x$, witnessed by $p'$ and $l'=l+1$, therefore $\model, z \models \psi$. If $\model, z \models \lnot \phi$, then by hypothesis there must be $k \in \{1, \ldots,l+1\}$ such that $\model, p'(k) \models \psi$. By definition of $p'$, it is not possible that $k \in \{1, \ldots,l\}$, thus $k = l+1$ and $\model, z \models \psi$. By this argument, we have $\model, x \models \phi \until \psi$ using the set $A_x$ to verify the definition of satisfaction.
\end{proof}

\begin{proof}(of \autoref{remark:duals})
\begin{enumerate}
\item 
$
\model,x \models \phi \ldualuntil \psi
\jstp \iff {\mbox{Definition of $\ldualuntil$}}
 \model,x \models \neg (\neg\psi \until \neg\phi)
 \stp \iff
 \model,x \not\models \neg\psi \until \neg\phi
\jstp \iff {\mbox{\autoref{thm:other-side-of-simple-discrete-paths-until}}}
\neg (\model,x\models \neg\psi \mbox{ and } \forall p : x \pto {}{}\forall l\in \nats: \model, p(l) \models \lnot \neg\psi \Rightarrow \exists k\in \{1, \ldots, l\}: \model,p(k) \models \neg\phi)
\stp \iff
\neg (\model,x\models \neg\psi \mbox{ and } \forall p : x \pto {}{}\forall l\in \nats:~ \neg(\model, p(l) \models \psi) \vee (\exists k\in \{1, \ldots, l\}: \model,p(k) \models \neg\phi))
\stp \iff
\model,x\models \psi \mbox{ or } \exists p : x \pto {}{}\exists l\in \nats:~ \model, p(l) \models \psi \wedge \neg(\exists k\in \{1, \ldots, l\}: \model,p(k) \models \neg\phi)
\stp \iff
\exists p : x \pto {}{}\exists l\in \nats:~ \model, p(l) \models \psi \wedge \forall k \in \{1, \ldots, l\}: \model,p(k) \models \phi
$
\item 
$
\model,x \models \leverywhere \phi
\jstp \iff {\mbox{Definition of $\leverywhere$}}
\model,x \models \phi \until \bot
\jstp \iff {\mbox{\autoref{thm:other-side-of-simple-discrete-paths-until}}}
\forall p : x \pto {}{}\forall l\in \nats: \model, p(l) \models \lnot \phi \Rightarrow \exists k \in \{1, \ldots, l\}: \model,p(k) \models \bot
\stp \iff
\forall p : x \pto {}{}\forall l\in \nats: \model, p(l) \models \phi
$  
\item 
$
\model,x \models \lsomewhere \phi
\jstp \iff {\mbox{Definition of $\lsomewhere$}}
\model,x \models \neg \leverywhere \neg\phi
\jstp \iff {\mbox{\autoref{remark:duals}~(2)}}
\neg (\forall p : x \pto {}{}\forall l\in \nats: \model, p(l) \models \neg\phi)
\stp \iff
\exists p : x \pto {}{}\exists l\in \nats: \model, p(l) \models \phi
$ 
\end{enumerate}
\end{proof}

\begin{proof}{\autoref{lemma:termination}}
Let $size(\Phi)$ be inductively defined as follow:
\begin{itemize}
\item $size(\top)=size(p)=1$
\item $size(\neg\Phi)=size(\lozenge\Phi)=1+size(\Phi)$
\item $size(\Phi\wedge\Psi)=size(\Phi\until \Psi)=1+size(\Phi)+size(\Psi)$
\end{itemize}

We prove by induction on the syntax of $\slcs$ formulae that for any quasi-discrete closure model $\model=((X,\closure_{R}),\eval)$,
and for any formula $\Phi$ function $\check$ terminates in at most $\mathcal{O}(size(\Phi)\cdot(|X|+|R|))$ steps. 

\medskip
\noindent
\emph{Base of Induction.} If $\Phi=\top$ or $\Phi=p$ the statement follows directly from the definition 
of $\check$. Indeed, in both these cases function $\check$ computes the final result in just $1$ step.

\medskip
\noindent
\emph{Inductive Hypothesis.} Let $\Phi_1$ and $\Phi_2$ be such that  for any 
quasi-discrete closure model $\model=((X,\closure_{R}),\eval)$,  function $\check(\model,\Phi_i)$, $i=1,2$, terminate
 in at most $\mathcal{O}(size(\Phi_i)\cdot(|X|+|R|))$ steps.

\medskip
\noindent
\emph{Inductive Step.} 

\begin{description}
\item[$\Phi=\neg\Phi_1$:] In this case function $\check$ first recursively computes the set $P=\check(\model, \Phi_1)$,
then returns $X-P$. By inductive hypothesis, the calculation of $P$ terminates in at most $\mathcal{O}(size(\Phi_1)\cdot(|X|+|R|))$ steps, while to compute $X-P$ we need $\mathcal{O}(|X|)$ steps. Hence, $\check(\model, \neg\Phi_1)$
terminates in at most  $\mathcal{O}(size(\Phi_1)\cdot(|X|+|R|))+\mathcal{O}(|X|)$. However:
\[
\begin{array}{cl}
 & \mathcal{O}(size(\Phi_1)\cdot(|X|+|R|))+\mathcal{O}(|X|) \\
\leq & \mathcal{O}(size(\Phi_1)\cdot(|X|+|R|))+\mathcal{O}(|X|+|R|) \\
= & \mathcal{O}((1+size(\Phi_1))\cdot(|X|+|R|)) \\
= & \mathcal{O}(size(\neg\Phi_1)\cdot(|X|+|R|))
\end{array}
\]

\item[$\Phi=\Phi_1\wedge\Phi_2$:] To compute  $P=\check(\model, \Phi_1\wedge \Phi_2)$ function $\check$
first computes $P=\check(\model, \Phi_1)$ and $Q=\check(\model, \Phi_2)$. Then the final result is 
obtained as $P\cap Q$. Like for the previous case, we have that the statement follows from inductive hypothesis and by using the fact that $P\cap Q$ can be computed in 
at most $\mathcal{O}(|X|)$. 

\item[$\Phi=\lozenge\Phi_1$:] In this case function $\check$ first computes, in at most $\mathcal{O}(size(\Phi_1)\cdot(|X|+|R|))$ steps, the set $P=\check(\model,\Phi_1)$. Then the final result is obtained as $\closure_{R}(P)$. Note that,
to compute  $\closure_{R}(P)$ one needs $\mathcal{O}(|X|+|R|)$ steps. According to 
\autoref{def:closure-operator-of-a-relation}, $\closure_{R}(P)$ is obtained as the union, 
computable in $\mathcal{O}(|X|)$ steps, of $P$ with $\{x \in X | \exists a\in P. (a,x)\in R\}$. The latter
can be computed in $\mathcal{O}(|R|)$ steps. Indeed, we need to consider all the \emph{edges} exiting from $P$. 
Hence, $\check(\model,\lozenge\Phi_1)$ terminates in a number of steps that is:
\[
\begin{array}{cl}
& \mathcal{O}(size(\Phi_1)\cdot(|X|+|R|))+\mathcal{O}(|X|)
+\mathcal{O}(|R|)\\
= & \mathcal{O}(size(\Phi_1)\cdot(|X|+|R|))+\mathcal{O}(|X|+|R|)\\
= & \mathcal{O}((1+size(\Phi_1))\cdot(|X|+|R|))\\
= & \mathcal{O}(size(\lozenge\Phi_1)\cdot(|X|+|R|))\\
\end{array}
\]

\item[$\Phi=\Phi_1\until\Phi_2$:] When $\Phi=\Phi_1\until\Phi_2$ function $\check$ recursively invokes function $\checkUntil$ that first computes the sets $P=\check(\model,\Phi_1)$, $Q=\check(\model,\Phi_2)$ and $T= \cboundary(P \cup Q)$.
By inductive hypothesis, the computations of $P$ and $Q$ terminate in at most $\mathcal{O}(size(\Phi_1)\cdot(|X|+|R|))$
and $\mathcal{O}(size(\Phi_2)\cdot(|X|+|R|))$ steps, respectively, while $T$ can be computed in $\mathcal{O}(|X|+|R|)$.
After that, the loop at the end of function $\checkUntil$ is executed. We can observe that:
\begin{itemize}
\item a point $x$ is added to $T$ only one time (i.e. if an element is removed from $T$, it is never reinserted in $T$);
\item all the points in $T$ are eventually removed from $T$;
\item each \emph{edge} in $\model$ is traversed at most one time.
\end{itemize}

The first two items, together with the fact that $\model$ is finite, guarantee that the loop terminates. The last item
guarantees that the loop terminates in at most $\mathcal{O}(|R|)$ steps\footnote{Note that this is the complexity for a DFS 
in a graph}. Summing up, the computation of $\check(\model,\Phi_1\until\Phi_2)$ terminates in at most
\[
\begin{array}{cl}
& \mathcal{O}(size(\Phi_1)\cdot(|X|+|R|))+\mathcal{O}(size(\Phi_2)\cdot(|X|+|R|))\\
& +\mathcal{O}(|X|+|R|)+\mathcal{O}(|R|)\\
= & \mathcal{O}((size(\Phi_1)+size(\Phi_2))\cdot(|X|+|R|))+\mathcal{O}(|X|+|R|)\\
= & \mathcal{O}((1+size(\Phi_1)+size(\Phi_2))\cdot(|X|+|R|))\\
= & \mathcal{O}(size(\Phi_1\until\Phi_2)\cdot(|X|+|R|))\\
\end{array}
\]

\end{description}

\end{proof}

\begin{proof}{\autoref{thm:sound-compl}}
The proof proceeds by induction on the syntax of $\slcs$ formulae. 

\medskip
\noindent
\emph{Base of Induction.} If $\Phi=\top$ or $\Phi=p$ the statement follows directly from the definition 
of function $\check$ and from \autoref{def:closure-semantics}. 

\medskip
\noindent
\emph{Inductive Hypothesis.} Let $\Phi_1$ and $\Phi_2$ be such that  for any finite
quasi-discrete closure model $\model=((X,\closure_{R}),\eval)$,  function $x\in \check(\model,\Phi_i)$ if and only if
$\model,x\models \Phi_i$, for $i=1,2$.

\medskip
\noindent
\emph{Inductive Step.} 

\begin{description}
\item[$\Phi=\neg\Phi_1$:] 

$x\in \check(\model, \neg\Phi_1)
\jstp \iff {\mbox{Definition of $\check$}}
x\not\in \check(\model, \Phi_1)
\jstp \iff {\mbox{Inductive Hypothesis}}
\model,x\not\models\Phi_1
\jstp \iff {\mbox{\autoref{def:closure-semantics}}}
\model,x\models\neg\Phi_1
$

\item[$\Phi=\Phi_1\wedge\Phi_2$:] 
$x\in \check(\model, \Phi_1\wedge\Phi_2)
\jstp \iff {\mbox{Definition of $\check$}}
x\in \check(\model, \Phi_1)\cap \check(\model, \Phi_2)
\stp \iff
x\in \check(\model, \Phi_1)\mbox{ and } x\in \check(\model, \Phi_2)
\jstp \iff {\mbox{Inductive Hypothesis}}
\model,x\models\Phi_1  \mbox{ and } 
\model,x\models\Phi_2 
\jstp \iff {\mbox{\autoref{def:closure-semantics}}}
\model,x\models\Phi_1\wedge\Phi_2
$

\item[$\Phi=\lozenge\Phi_1$:] 
$x\in \check(\lozenge\Phi_1)
\jstp \iff {\mbox{Definition of $\check$}}
x\in \closure_{R}(\check(\model, \Phi_1))
\jstp \iff {\mbox{Definition of $\closure_{R}$}}
\exists A\subseteq \check(\model, \Phi_1): x\in \closure_{R}(A)
\jstp \iff {\mbox{Inductive Hypothesis}}
\exists A\subseteq X. \forall y\in A. \model,y,\models \Phi_i\mbox{ and } x\in \closure_{R}(A)
\jstp \iff {\mbox{\autoref{def:closure-semantics}}}
\model,x\models \lozenge\Phi_1
$

\item[$\Phi=\Phi_1\until\Phi_2$:] We prove that $x\in \checkUntil(\model,\Phi_1,\Phi_2)$ if and 
only if $\model,x\models \Phi_1\until\Phi_2$. 
Function $\checkUntil$ takes as parameters a model $\model$ and two $\slcs$ formulas $\Phi_1$ and $\Phi_2$ and 
computes the set of points in $\model$ satisfying $\Phi_1 \until \Phi_2$ by removing from $V=\check(\model,\Phi_1)$
all the \emph{bad} points. 

A point is \emph{bad} if it can \emph{reach} a point satisfying $\neg\Phi_1$ 
without passing through a point satisfying $\Phi_2$.
Let $Q=\check(\model,\Phi_2)$ be the set of points in $\model$ satisfying $\Phi_2$. 
To identify the \emph{bad} points in $V$ the function \checkUntil performs a \emph{backward search} from 
$T=\cboundary(V\cup Q)$.  Note that any \emph{path exiting} from $V\cup Q$ has to pass through points
in $T$. Moreover, the latter only contains points that satisfy neither $\Phi_1$ nor $\Phi_2$, by definition.
Until $T$ is empty, function \checkUntil first picks all the elements $x$ in $T$ and then removes from $V$ 
the set of (bad) points $N$ that are in $V-Q$ and that can reach  $x$ in \emph{one step}.  
%
%
At the end of each iteration the set $T$ contains the set of \emph{bad} points discovered in the last
iteration.
The proof proceeds in two steps. The first step guarantees that if $x$ does not satisfy $\Phi_1\until \Phi_2$,
then $x$ is eventually removed from $V$. The second step shows that if $x$ is removed from $V$
then $x$ does not satisfy $\Phi_1\until \Phi_2$.

Note that, by Inductive Hypothesis, we have that: 

\begin{equation}
\label{eq:assumption1}
x\in V=\check(\model,\Phi_1) \Leftrightarrow \model,x\models \Phi_1
\end{equation}

\begin{equation}
\label{eq:assumption2}
x\in Q=\check(\model,\Phi_2) \Leftrightarrow \model,x\models \Phi_2
\end{equation}

For each $x\in X$ we let:
\[
\mathcal{I}_{x} = \{ i\in \mathbb{N} | \exists p:x\pto{}{}. \model,p[i] \models\neg\Phi_1\wedge\forall j\in \{ 1,\ldots, i \}. \model,p[j]\models\neg\Phi_2 \}
\]
Note that, directly from \autoref{thm:other-side-of-simple-discrete-paths-until}, we have that
$\model,x\models \Phi_1\until \Phi_2$ if and only if $\model,x\models \Phi_1$ and $\mathcal{I}_{x}=\emptyset$.

First we prove that 
if $\mathcal{I}_{x}\not=\emptyset$ and $\model,x\models \Phi_1$, then $x$ is removed from 
$V$ at iteration $i=\min \mathcal{I}_{x}$. This guarantees that if $x$ does not satisfy $\Phi_1\until \Phi_2$,
then $x$ is eventually removed from $V$. The proof of this result proceeds by induction on $i$:

\begin{description}
\item[Base of Induction:] Let $x\in X$ such that $\model,x\models \Phi_1$, $\mathcal{I}_{x}\not=\emptyset$ and 
$\min \mathcal{I}_{x}=1$. 
Since $\min \mathcal{I}_{x}=1$, we have that there exists $p:x\pto{}{}$ such 
that $\model,p[1]\models \neg\Phi_1$ and $\model,p[1]\models \neg\Phi_2$. 
By definition of paths, we also have that $x=p[0]$ and $(x,p[1])\in R$. This implies that $p[1]\in \cboundary(V\cup Q)$
and $x\in pre(p[1])$. By definition of function $\checkUntil$ we have that $p[1]$ is in $T$ and 
$x$ is removed from $V$ during the first iteration.
Note that $x$ will be added to $T$ only if it does not satisfy $\Phi_2$ (i.e. if $x\not\in Q$).  

\item[Inductive Hypothesis:] For each $x\in X$ be such that $\model,x\models \Phi_1$, $\mathcal{I}_{x}\not=\emptyset$ and 
$\min \mathcal{I}_{x}=k$, $x$ is removed from $V$ at iteration $k$.

\item[Inductive Step:] Let $x\in X$ be such that $\model,x\models \Phi_1$, $\mathcal{I}_{x}\not=\emptyset$ and 
$\min \mathcal{I}_{x}=k+1$. 
If $\min \mathcal{I}_{x}=k+1$ then there exists $p:x\pto{}{}$ such that $\model,p[k+1]\models \neg\Phi_1$ and 
for each $j\in \{1,\ldots,k+1\}$ $\model,p[j]\models \neg\Phi_2$. 
We have also that $\model, p[1] \models \Phi_1$ (otherwise $\min \mathcal{I}_{x}=1$) and 
$\min \mathcal{I}_{p[1]}=k$  (otherwise $\min \mathcal{I}_{x}\not=k+1$).
By inductive hypothesis we have that $p[1]$ is removed from $V$ at iteration $k$. However,
since $\model,p[1]\models\neg \Phi_2$ we have that $p[1]\not \in Q$ and $p[1]$ is in the set $T$ 
at the beginning of iteration $k+1$. 
This implies that $x=p[0]$ is removed from $V$ at iteration $k+1$, since $x\in pre(p[1])$.
\end{description}

\medskip
We now prove that if $x$ is removed from $V$ at iteration $i$, then $\mathcal{I}_{x}\not=\emptyset$ and $i=\min \mathcal{I}_{x}$. 
This ensures that if $x$ is removed from $V$ then $x$ does not satisfy $\Phi_1\until \Phi_2$.
We proceed by induction on the number of iterations $i$:

\begin{description}
\item[Base of Induction:] If $x\in V$ is removed in the first iteration we have that there exists a point 
$y\in \cboundary(V\cup Q)$ such that $(x,y)\in R$. From \autoref{eq:assumption1} 
and \autoref{eq:assumption2} we have that $\model, x \models \Phi_1$ while $\model,y\models\neg\Phi_1\wedge\neg\Phi_2$. This implies that there exists a path $p:x\pto{}{}$ such that $p[1]=y$ and $1=\min \mathcal{I}_{x}$. 

\item[Inductive Hypothesis:] For each point $x\in V$,  if $x$ is removed from $V$ at iteration $i\leq k$, then 
$\mathcal{I}_{x}\not=\emptyset$ and $i=\min \mathcal{I}_{x}$. 

\item[Inductive Step:] Let $x\in V$ be removed at iteration $k+1$.  This implies that after $k$ iterations, 
there exists a point $y$ in $T$ such that $(x,y)\in R$. This implies that $y$ has been removed
from $V$ at iteration $k$ and, by inductive hypothesis, $\mathcal{I}_{y}\not=\emptyset$ and 
$k=\min \mathcal{I}_{y}$. Hence, there exists a path $p:y\pto{}{}$ such that $\model,p[k]\models \neg\Phi_1$ and 
for each $j\in \{1,\ldots,k\}$ $\model,p[j]\models \neg\Phi_2$.
Moreover, since $y\in T$, we have also that $y\not\in Q$ and, from \autoref{eq:assumption2}, 
$\model,y\models\neg\Phi_2$. We can consider the path $p':x\pto{}{}$ such that, for each $j$, $p'[0]=x$ and
$p'[j+1]=p[j]$. We have that $\model,p'[k+1]\models\neg\Phi_1$ and for each $j\in \{1,\ldots,k+1\}$, 
$\model,p'[j]\models\neg\Phi_2$. Hence $\mathcal{I}_{x}\not=\emptyset$ and $k+1=\min \mathcal{I}_{x}$
(otherwise $x$ should be removed from $V$ in a previous iteration).

\end{description}

\end{description}

\end{proof}